Synthesis, Characterization, and Modeling of Naphthyl-Terminated *sp* Carbon Chains:

Dinaphthylpolyynes

*Franco Cataldo\** [1,2], *Luca Ravagnan\** [3,4], *Eugenio Cinquanta*[4,5], *Ivano Eligio Castelli*[3,6],

*Nicola Manini*[3,7], *Giovanni Onida*[3,7], *Paolo Milani*[3,4]

Actinium Chemical Research, Via Casilina 1626/A, I-00133 Rome, Italy,

Istituto Nazionale di Astrofisica. Osservatorio Astrofisica di Catania, Via S. Sofia 78, I-95123 Catania,

Italy,

Dipartimento di Fisica, Università degli Studi di Milano, Via Celoria 16, I-20133 Milano, Italy,

CIMAINA, Via Celoria 16, I-20133 Milano, Italy,

Dipartimento di Scienza dei Materiali, Università degli Studi di Milano-Bicocca, Via Cozzi 53, I-20125

Milano, Italy, and

European Theoretical Spectroscopy Facility (ETSF), Via Celoria 16, 20133 Milano, Italy.

---

\* To whom correspondence should be addressed. E-mail: franco.cataldo@fastwebnet.it (F.C.);

luca.ravagnan@mi.infn.it (L.R.).

[1] Istituto Nazionale di Astrofisica.

[2] Actinium Chemical Research.

[3] Dipartimento di Fisica, Università degli Studi di Milano.

[4] CIMAINA.

[5] Dipartimento di Scienza dei Materiali, Università degli Studi di Milano-Bicocca.

[6] Present address: Center for Atomic-scale Materials Design (CAMD), Department of Physics,

Technical University of Denmark, DK-2800 Kongens Lyngby, Denmark.

**ABSTRACT**: We report a combined study on the synthesis, spectroscopic characterization and theoretical modelling of a series of α,ω-dinaphthylpolyynes. We synthesized this family of naphtyl-terminated *sp* carbon chains by reacting diiodoacetylene and 1-ethynylnaphthalene under the Cadiot-Chodkiewicz reaction conditions. By means of liquid chromatography (HPLC), we separated the products and recorded their electronic absorption spectra, which enabled us to identify the complete series of dinaphthylpolyynes Ar-C$_{2n}$-Ar (with Ar = naphthyl group and *n* = number of acetilenic units) with *n* ranging from 2 to 6. The longest wavelength transition (LWT) in the electronic spectra of the dinaphthylpolyynes red shifts linearly with *n* away from the LWT of the bare termination. This result is also supported by DFT-LDA simulations. Finally, we probed the stability of the dinaphthylpolyynes in a solid-state precipitate by Fourier-transform infrared spectroscopy and by differential scanning calorimetry (DSC).

---

[7] European Theoretical Spectroscopy Facility.

1. **Introduction**

During the past years the study of *sp* carbon chains (*sp*CCs, also known as acetylenic arrays) has attracted the interest of a number of research group from a multitude of disciplinary areas, ranging from organic chemistry, non linear optics, solid state physics, and astrophysics. Although bare *sp*CCs are extremely unstable and reactive species [1], it has been shown that their stability is remarkably enhanced when their reactive terminations are bond to chemical groups [2-4] or to carbon nanostructures [5,6].

For instance, chemists were able to synthesize *sp*CCs which are stable in solution or even in the solid phase, by terminating the chains with organic group [3,4]. This class of compounds, often referred to as α,ω-diarylpolyynes (DAPs), have interesting and promising physical properties. For instance they exhibit a HOMO-LUMO gap strongly influenced by the number of *sp* carbon atoms constituting the chain [7], and their third-order nonlinear optical properties and their nonresonant molecular second hyperpolarizabilities ($\gamma$) have been reported to increase as a function of the chain length [8,9]. Furthermore, it has been shown that the fluorescence, phosphorescence and absorption spectra of DAPs can be modulated by applying an external electric field [10]. Finally, unusual photophysical and spectroscopic properties have been reported for DAPs, and they are considered promising candidates as materials for the fabrication of semiconductive thin film transistor structures [11], as the active semiconductor material in light-emitting diodes in electroluminescent devices [12,13] and in organic solar cells [14].

On the other hand, solid state physicists are studying the possibility to produce carbon solids containing *sp*CCs (*sp-sp$^2$* carbon) since the 1960's [15]. Although the feasibility to obtain *sp-sp$^2$* carbon solids has been shown for some years [16,17], recent experimental observation by HR-TEM of *sp*CCs forming during the irradiation of graphene planes has strongly increased the interest on this field [18]. *sp*CCs are nowadays considered promising structures for nano-electronic applications [19,20] since they could be used as molecular conductors bridging graphene nanogap devices. First potential applications have been already demonstrated for the realization of non-volatile memories, as their use in two-terminal atomic-scale switches [20]. Furthermore, *sp*CCs are expected to have interesting rectifying performances [21], to become spin-polarized depending to their axial strain [5], and to be effective as spin-filters and spin-valves [22].

Finally *sp*CCs are of great relevance for astrophysics in the framework of the study of the Diffuse Interstellar Bands (DIBs). DIBs include a large number of absorption lines in the visible-near IR region of the electromagnetic spectrum that are superposed on the interstellar extinction curve and whose identification remains one of the oldest mysteries in stellar spectroscopy [23]. By comparing laboratory spectra of complex molecules with the DIBs lines, several molecules were identified as constituent of the interstellar medium [23,24]. Indeed, both non-terminated *sp*CCs and polycyclic aromatic hydrocarbons (PAHs), as for example naphthalene, have been identified separately [25-27], however *sp*CCs terminated by PAHs (e.g. DAPs) have not been considered to date.

In the present paper we present a combined study on the synthesis, spectroscopic characterization and theoretical modelling of a series of *sp*CCs terminated by naphtyl groups (dinaphthylpolyynes, in short represented as Ar-C$_{2n}$-Ar, with Ar = naphthyl group and *n* = number of acetylenic units of the *sp*CC). These molecules, already synthesized in the 1960's and 1970's by the group of M. Nakagawa [28-30], represent a family of chemically stabilized DAPs which are of high interest for several reasons: from one side, they are potentially useful in optoelectronic applications and are promising building blocks for the production of *sp-sp$^2$* carbon systems for nano-electronic devices; from the other side, they constitute the simplest examples of *sp*CCs terminated by a PAH that could be identified in the interstellar medium (naphthalene is indeed one of the simpler and most abundant PAHs). Here we present a new route for the synthesis of this class of complexes which enables, by a remarkably simpler chemical approach than in Ref. [28-30], to produce at the same time a solution of dinaphthylpolyynes Ar-C$_{2n}$-Ar with 2≤n≤6.

## 2. Experimental Section

### 2.1 Materials and Equipment

All the solvents and reagents (as for instance 1-ethynylnaphthalene) mentioned in the present work were obtained from Sigma Aldrich. Diiodoacetylene was prepared according to the method reported in Ref. [31]. All the equipment and methods for analysis and characterization of the reaction products are reported in Ref. [32].

We performed the HPLC analysis using an Agilent Technologies 1100 station with $C_8$ column (using acetonitrile/water 80/20 as mobile phase at a flow rate of 1.0 mL/min) and equipped with a diode array detector (used for recording the UV-vis absorption spectra of the separated molecules). UV-vis absorption spectra were also acquired on the solution and thin films by a JASCO-7850 spectrophotometer. FT-IR measurements were carried out on in transmittance mode on an IR300 spectrometer from Thermo-Fisher Corp. The differential scanning calorimetric (DSC) measurements were performed on a DSC-1 Star System from Mettler-Toledo (heating rate of 10°C/min under $N_2$ flow).

2.2 Synthesis of Dinaphthylpolyynes

We synthesized the dinaphthylpolyynes by reacting copper(I)ethynylnaphthalide with diiodoacetylene under the Cadiot-Chodkiewicz reaction conditions [32,33].

In a first stage we synthesized the copper salt of ethynylnaphthalene by dissolving Cu(I) chloride (1.0 g) in 30 mL of aqueous ammonia (30%) together with 0.5 g of hydroxylamine hydrochloride ($NH_2OH \cdot HCl$). 1-Ethynylnaphthalene (1.5 mL) was then added to this solution under stirring. We collected a dark yellow precipitate of copper(I)ethynylnaphthalide by filtration in considerable yield (1.8 g).

After that about 6.0 g of I-C≡C-I (Warning! Acetylene diiodide is an irritant and poisonous) was dissolved/suspended in 90 mL of decalin, it was transferred in a 500 mL round bottomed flask and was stirred with 100 mL of distilled water. Copper(I)-ethynylnaphthalide (1.8 g) was then added to the reaction mixture together with 25 mL of tetrahydrofuran, 50 mL aqueous $NH_3$ 30% and 30 mL of N,N',N,N'-tetramethylethylenediamine (TMEDA). The mixture was stirred at room temperature for 2 days and after this process the decalin layer became deep orange. The reaction mixture was filtered with the aid of an aspirator to separate the black residue formed. The filtered solution consisted of two layers. The organic layer was separated from the aqueous layer by means of a separatory funnel. The resulting crude solution contained dinaphthylpolyynes dissolved in decalin, together with a significant amount of free unreacted ethynylnaphthalene and other by-products. These latter were removed from the crude solution (as a dark red precipitate) by shaking the solution with 4.0 g of CuCl dissolved in 100 mL of aqueous $NH_3$ 30% together with 3.0 g of $NH_2OH \cdot HCl$.

The resulting orange solution was then characterized by HPLC, UV-vis, FT-IR and DSC.

2.3 DFT Simulations

We simulated the synthesized structures using standard DFT-LSDA (density functional theory in the local spin density approximation), based on a plane-waves basis, and pseudopotentials [34]. We relaxed all atomic positions until the largest residual force was <8 pN. The ensuing structures and electronic Kohn-Shan orbitals are depicted using a standard visualization tool [35].

**3. Results and Discussion**

3.1 Synthesis and Identification of Dinaphthylpolyynes

Previous work [32] demonstrated the great versatility of diiodoacetylene ($C_2I_2$) as a building block for the synthesis of symmetrically terminated *sp*CCs through the Cadiot-Chodkiewicz reaction. The synthetic approach adopted previously with copper phenylacetylide and $C_2I_2$ has been extended here to the case of the copper salt of 1-ethynylnaphthalene. As shown in Scheme 1, the resulting *sp*CCs are thus terminated by the bulkier naphthalene groups.

Figure 1 reports the separation by elution through the HPLC column of the products formed in the Cadiot-Chodkiewicz reaction between diiodoacetylene and 1-ethynylnaphthalene. The presence of well distinct peaks at different retention times demonstrates the formation of products with different molecular weights, in agreement with the formation of dinaphthyl terminated *sp*CCs of different length (according to the Scheme 1). The formation of *sp*CCs having more than three acetylenic units suggests the formation of the labile asymmetric intermediate Cu-C≡C-I during the coupling reaction [32]. Without the formation of this intermediate, the production of *sp*CCs longer than three acetylenic units would be unjustified.

By using a low flow rate for the mobile phase in the HPLC it was possible to observe the regular elution of all dinaphthylpolyynes, although a very long retention time was observed for the products with higher molecular weight. Figure 1 shows the retention times of the dinaphthylpolyynes in the HPLC column (as a function of the number of acetylenic units) in comparison to the diphenylpolynes ones [32]. The retention times of both series of molecules can be fitted by an exponential law of the type (see inset of Figure 1):

$R_t = A e^{B \cdot n}$     (eq.1)

where $R_t$ is the retention time in the column, $n$ is the number of acetylenic units of the *sp*CC, and $A$ and $B$ are two parameters whose values are reported in Table 1.

| *sp*CCs series | $A$ [min] | $B$ |
|---|---|---|
| dinaphthylpolyynes | 3.0 ± 0.3 | 0.53 ± 0.02 |
| diphenylpolyynes | 1.4 ± 0.1 | 0.51 ± 0.02 |

Table 1. Fitting parameters for the exponential law describing the evolution of $R_t$ with $n$ (eq. 1).

Remarkably, the longer retention times of dinaphthylpolyynes in comparison to those of diphenylpolyynes, (ascribable to their bulkier end groups) seems to affect only the parameter $A$ of the exponential fit, which is almost double in the case of dinaphthylpolyynes in comparison to the value obtained for diphenylpolyynes. In contrast, the exponents $B$ obtained for the two series of chains are comparable.

3.2 Electronic Absorption Spectroscopy

By using the diode array detector interfaced to the column of the HPLC, we could record the electronic absorption spectrum of each molecular species contained in the dinaphthylpolyynes mixture, i.e. for each peak in Figure 1. Figure 2 shows the resulting spectra. The presence of a regular shift for the position of the absorption peaks as the molecular weight of the species increases (i.e., the retention time increases) supports the assignment to diphenylpolyynes of increasing length. Indeed, it is well known that the position of the electronic absorption peaks of *sp*CCs is correlated to the number of acetylene units [28-30,7]. The exact attribution of the different spectra to the corresponding dinaphthylpolyyne is furthermore possible by comparing the absorption spectra with the one obtained by the group of M. Nakagawa [28-30]. Remarkably, the position of the peak at longest wavelength (longest wavelength transition, LWT) of our spectra matches the values in ref. [28-30], corresponding to dinaphthylpolyynes with *n*=2-6, within a shift of about 3 nm. The presence of this shift has to be ascribed to the different solvent used in our case (decalin) with respect to the one used in the past (tetrahydrofuran) [29].

Considering the different spectra in detail, they are characterized by a complex structure, in which two main regions can be identified. The region at higher wavelength (labeled as Reg. A in Figure 2a) is related to the electronic transitions between the highest occupied molecular orbital (HOMO) and the lowest unoccupied molecular orbital (LUMO); the region at lower wavelength (labeled as Reg. B in Figure 2a) is related to transitions between other molecular orbitals (such as, for instance, the transitions between the HOMO and LUMO+1 or the HOMO-1 and LUMO). In region A in particular, a multipeak structure can be identified that can be fitted as the superposition of several Gaussian components. The presence of several peaks can be attributed to the Franck-Condon shifts of the main HOMO-LUMO transition [28-30,7], i.e. the simultaneous occurrence of the electronic transition and the excitation of vibrational modes of the molecule. In this framework, the LWT peak of the electronic absorption spectrum corresponds to the HOMO-LUMO transition without the excitation of the molecule to a new vibrational level, while the most intense transition (MIT) is the transition for which the starting and final wave functions overlap more significantly (thus, the transition has a higher probability to occur). As shown in Figure 2a dinaphthylpolyynes have MITs that are always distinct by the LWT. This is a characteristic of *sp*CCs terminated by bulky groups, and indeed has been observed also for diphenylpolyynes [32] and for *sp*CCs terminated by dendridic organic groups [3]. Instead, in the case of simple polyynes, the LWT and the MIT coincide [2].

Figure 2b shows the same spectra as Figure 2a (limited to Reg. A), where the abscissa is expressed as energy shift from the LWT. This means that the wavelength scale of each spectrum in Figure 2a was first converted in energy scale (in cm$^{-1}$ units), and then, the energy of the LWT peak was subtracted from it; as a result, in Figure 2b the LWTs of all spectra are centered at 0 cm$^{-1}$. Thanks to this representation, it is possible to observe that the energy shift between the peaks in region A is on the order of 1900-2000 cm$^{-1}$, which is consistent with the typical main vibrational modes of *sp*CCs [36]. Furthermore, the value of this shift decreases as the chain length of the dinaphthylpolyyne increases: this again is consistent with the behavior of the vibrational modes of *sp*CCs that are known to be softer (i.e., characterized by a lower energy) when the chain length is larger [36]. As a matter of fact, this behaviour was also observed in Ref. [29], even if the Franck-Condon shifts measured in our case are slightly lower. Like for the shift of the LWT, this slight deviation can be related to the different solvent.

Figure 3 shows the evolution of the LWT of the electronic absorption spectra of the dinaphthylpolyynes as a function of the number of acetylenic units ($n$) in the chain, compared to the same evolution for diphenylpolyynes [32] and hydrogen-terminated *sp*CCs (i.e., polyynes) [2]. In all cases, a linear relation emerges between the LWT and the number of acetylenic units [7], as described by the following equation:

$LWT = \alpha \cdot n + LWT_0$ (eq.2)

where $LWT_0$ is the extrapolated LWT for a chain of "null" length and $\alpha$ is the slope. Although other groups have suggested that LWT for dinaphthylpolyynes evolves as $LWT = \alpha \cdot n^{1/2} + LWT_0$ [28-30] our data are more consistent with a linear dependence with $n$. The fitting parameters of $LWT_0$ and $\alpha$ obtained for the three series of *sp*CCs in Figure 3 are reported in Table 2.

| *sp*CCs series | $\alpha$ | $LWT_0$ |
| --- | --- | --- |

|  | [nm/$C_2$ unit] | [nm] |
|---|---|---|
| Dinaphthylpolyynes | 25.7 ± 0.5 | 319.8 ± 2.2 |
| Diphenylpolyynes | 35.3 ± 0.8 | 256 ± 3 |
| Polyynes | 23.0 ± 0.5 | 133 ± 3 |

Table 2. Fitting parameters for the linear relation describing the evolution of LWT with $n$ (eq. 2).

Remarkably, the value of $LWT_0$ for dinaphthylpolyynes at 319.8 is very close to LWT of the the B-band of naphthalene at 313 nm [37-39] as well as to the LWT of binaphthyl at 315 nm [37-39]. Similarly, the value of $LWT_0$ for diphenylpolyynes at 256 nm is close to LWT of the B band of benzene at 267 nm [37] and is very close to the LWT for biphenyl at about 252 nm [37].

As suggested in Ref. [30], these observations indicate that the intercepts at $n=0$ correspond to the LWT absorption of the end groups, while the addition of the acetylenic unit causes a red shift in the LWT transition.

In order to better understand the observed behavior, we carried out DFT-LSDA simulations of the geometric and electronic structure of a free-standing naphthalene molecule and of dinaphthylpolyynes with different chain lengths. Naphthalene molecules exibit a HOMO-LUMO gap larger than that of a dinaphthylpolyynes of any length. Figure 4 (left panel) shows the HOMO and LUMO wave functions obtained for Ar-$C_4$-Ar, Ar-$C_8$-Ar and Ar-$C_{12}$-Ar. As can be observed, the HOMO and LUMO orbitals of the *sp*CCs are hybridized with those of the terminations since their wave functions extend out from the *sp*-chain region in the naphthyl region. Nevertheless, as the chain length increases, the HOMO and LUMO orbitals become more and more localized on the sole chain backbone, indicating a diminishing hybridization. At the same time, as the chain length increases, the computed HOMO-LUMO gap decreases, and consequently the theoretical value for the absorption wavelength ($LWT_{DFT}$, proportional to the inverse of the computed HOMO-LUMO gaps) increases. Figure 4 (right panel) shows the evolution of $LWT_{DFT}$ with $n$. Although the computed values for the $LWT_{DFT}$ do not match the experimental values for the LWT in Figure 3, the overall trend is in very good agreement. The simulations reproduce the linear relation between $LWT_{DFT}$ and $n$ observed experimentally, and the intercept at $n=0$ is not far from to the simulated LWT$_{DFT}$ for naphthalene, given the expected underestimation of the HOMO-LUMO gap of DFT-LSDA.

It is also relevant to note that in the case of polyynes the $\pi$ and $\pi^*$ orbitals (i.e., the HOMO and LUMO orbitals) are completely localized on the carbon chain axis, and therefore, little or no hybridization of those orbitals occurs with the termination groups. Nevertheless, the same linear relation between LWT and $n$ is observed. This indicates that the slope $\alpha$ of the linear relations in eq. 2 is not simply a measure of the hybridization between the $\pi$ electron state of the chains and $\pi$ electron state of the end groups but reflects a gap reduction intrinsic of the carbyne chain. The decrease in chain end-group hybridization seems in fact a direct consequence of this intrinsic gap reduction, with a corresponding movement "off-resonance" of the chain HOMO and LUMO levels relative to those of the end groups.

### 3.3 Relative Yield of Dinaphthylpolyynes

We compute the relative yield $C(n)$ of dinaphthylpolyynes of different lengths from the intensity of the most intense transition in the electronic absorption spectra (i.e., the MIT in Figure 2). We take advantage of the molar extinction coefficients $\varepsilon(n)$ previously reported [29]. We use the following relation:

$$C(n) = \frac{\dfrac{I_{MIT}(n)}{\varepsilon(n)} \cdot \dfrac{A(n,\lambda)}{I(n,\lambda)}}{\sum_{i=2}^{6} \dfrac{I_{MIT}(i)}{\varepsilon(i)} \cdot \dfrac{A(i,\lambda)}{I(i,\lambda)}} \qquad (eq.3)$$

where $I_{MIT}(n)$ is the intensity (i.e., the absorbance) of the MIT of the absorption spectrum for the dinaphthylpolyyne with $n$ acetylenic units (see Figure 2), $I(n,\lambda)$ is the value of absorbance measured in the same spectrum at the wavelength $\lambda$ and $A(n,\lambda)$ is the time-integrated intensity of the corresponding peak in the HPLC spectrum acquired using the wavelength $\lambda$. Table 3 reports the average of the values of $C(n)$ obtained by applying eq. 3 to two different HPLC spectra, acquired at $\lambda=350$ nm (see Figure 1a) and $\lambda=250$ nm (not shown) respectively.

| *n* | dinaphthylpolyyne formulas | *C(n)* |
|---|---|---|
| 2 | Ar-C$_4$-Ar | 47 ± 4 % |
| 3 | Ar-C$_6$-Ar | 19.7 ± 0.6 % |
| 4 | Ar-C$_8$-Ar | 18.2 ± 2.2 % |
| 5 | Ar-C$_{10}$-Ar | 11.3 ± 1.2 % |
| 6 | Ar-C$_{12}$-Ar | 4.1 ± 0.3 % |

Table 3. Relative molar concentration of the dinaphthylpolyynes (with 2≤*n*≤6) in the solution.

Figure 5 shows the relative molar concentration of the dinaphthylpolyynes in comparison to relative molar concentration of the diphenylpolyynes series produced using the same synthetic approach [32]. As can be observed, for any chain length larger than *n* = 2 the concentration of dinaphthylpolyynes exceeds that of the diphenylpolyynes. Only at *n* = 2 the concentration of diphenylpolyynes is higher than that of dinaphthylpolyynes. These data suggest that our synthetic approach favors longer chains when bulkier terminations are used as end groups (i.e., naphthyl groups) and shorter chains when the lighter phenyl groups are employed. Nevertheless, in both cases, the shortest chains are predominant, and the abundance of the species with longer chains decreases steadily with the increase of *n*, as normally expected in the synthesis of mixtures of polyynes [2].

In the calculation of the distribution of the dinaphthylpolyynes produced by the Cadiot-Chodkiewicz synthesis, we have neglected the presence of the residual 1-ethynyl-naphthalene which is not detected in the chromatogram shown in Figure 1a that was recorded at 350 nm but was present and detected in the chromatogram recorded at 250 nm (not shown).

3.4 Dinaphthylpolyynes in the solid phase

As previously discussed, *sp*CCs are interesting candidates for the development of a new generation of nanoelectronic devices, where they could be used as molecular conductors or spin filters and spin valves [19,20,22]. In order to produce similar devices, it is necessary to introduce *sp*CCs in solid-state systems and to ensure their stability in time. Unfortunately, *sp*CCs are notoriously unstable structures, and, in particular, if they are terminated simply by hydrogen or nitrogen atoms, they can be prepared only as very diluted solutions; if they are deposited as thin films, they undergo fast decay [2]. Furthermore, *sp*CCs are known to be highly reactive with oxygen [16,40], making the production of devices containing *sp*CCs quite challenging. Nevertheless, several groups were able to demonstrate that *sp*CCs terminated by bulky end groups have remarkably higher stability [3,4], so that in some cases even their crystallization was achieved [4]. In order to study the possibility to use dinaphthylpolyynes in solid-state systems, we have deposited droplets of the solution of dinaphthylpolyynes in decalin on hard substrates (made by CaF, ZnSe, and KBr depending on the characterization carried out), and we let the solvent evaporate by placing the sample in a dryer. The resulting film was then characterized by UV-vis and FT-IR spectroscopies and by DSC.

In Figure 6, we compare the UV-vis spectra of the whole solution of diphenylpolyynes in decalin (not separated by the HPLC column) and of the film obtained by the same solution after the evaporation of the solvent. As it is possible to observe, the complex structure present in the solution's spectrum (corresponding to the overlap of the separate spectra of Figure 2a) is radically lost in the dried film, where three broad peaks (indicated by arrows in the figure) are still distinguishable on top of an amorphous band. This indicates that during the evaporation of the solvent, the dinaphthylpolyynes undergo chemical reactions that do not completely destroy the molecules, as indicated by the survival of spectral features in the UV-vis spectrum.

To gather more detailed information on the structure of the "polymerized" polyynes, we carried out FT-IR measurements on the samples. Figure 7 shows the FT-IR spectrum of the dried dinaphthylpolyynes film measured in ambient conditions (i.e., the sample has been exposed to air) as compared to the spectrum of pure naphthalene [41] and pure 1-ethynylnaphthalene. The figure also shows the FT-IR spectrum of the dark precipitate depositing on the bottom of the sealed flask containing the dinaphthylpolyynes solution after months of the solution being stored at room temperature.

As can be observed, the FT-IR spectra of both 1-ethynylnaphthalene and of the dried dinaphthylpolyynes exibit the vibrational modes typical of naphthalene (as well as of the naphthyl group). For instance, we can identify in all the spectra the CH stretching band at about 3060 cm$^{-1}$, the in-phase –CH– out-of-plane bending band at about 780 cm$^{-1}$ (CH wagging) and a series of C=C stretching bands (at about 1590, 1506 and 1390 cm$^{-1}$) [42]. 1-ethynilnaphthalene is furthermore characterized by the triple bond ≡C-H stretching band at 3282 cm$^{-1}$ ($\nu_{≡C-H}$), by the bending of the ≡C-H at about 600 cm$^{-1}$ ($\delta_{≡C-H}$) and by the acetylenic stretching band at 2100 cm$^{-1}$ ($\nu_{C≡C}$) [43]. Remarkably, both $\nu_{≡C-H}$ and $\delta_{≡C-H}$ are no any longer detectable in the spectrum of the dried dinaphthylpolyynes, and the $\nu_{C≡C}$ of 1-ethynylnaphthalene at 2100 cm$^{-1}$ is replaced by a complex and relatively more intense band at 2194 cm$^{-1}$ with a series of sub-bands (at 2178, 2153 and 2125 cm$^{-1}$; see left panel in Figure 7). Indeed, the position of those bands is consistent with the C≡C stretching modes for isolated dinaphthylpolyynes [29].

Thes observations indicates on one hand that the interaction of the dinaphthylpolyynes occurring during their drying (as indicated by the modification of the UV-vis spectrum) is not altering significantly the dinaphthylpolyyne structure. In the FT-IR spectrum, we can indeed identify clearly the vibrational modes of both the naphthyl terminations and the C≡C stretching modes of the *sp*CCs, whose peaks do not show any significant broadening in comparison to the case of naphthalene and 1-ethynilnaphthalene.

On the other side, the complete absence in the dried dinaphthylpolyynes spectrum of the $\nu_{≡C-H}$ and $\delta_{≡C-H}$ vibrational modes can be interpreted as the indication that the Cadiot-Chodkiewicz reaction of 1-ethynylnaphathalene with diodoacetylene was complete and that the reaction products are essentially constituted by dinaphthylpolyynes. The residual, unreacted 1-ethynylnaphthalene detected in the HPLC analysis of the dinaphthylpolyynes (using the 250 nm wavelength) is thus a minor component not detectable by FT-IR spectroscopy.

Finally, it is significant to consider the FT-IR spectrum of the dark precipitate in Figure 7. The spectrum exibits four main broad bands, whose positions are still consistent with the main vibrational contributions observed for the dried dinaphthylpolyynes. We can distinguish two broad bands at about 3000 cm$^{-1}$ and 750-800 cm$^{-1}$, consistent with the CH stretching and CH wagging of the dinaphthyl terminations, respectively [42], an intense band in the 1300-1600 cm$^{-1}$ region that can be attributed to C=C stretching modes, and a band in the 2000-2200 cm$^{-1}$ region, related to C≡C stretching modes. This indicates that the dark precipitate is the result of the polymerization of the dinaphthylpolyynes in the solution. Remarkably anyway, the polymerization process is only capable of inducing a broadening of the vibrational bands, but it does not cause a significant depletion of the *sp*CCs. Furthermore, the obtained spectrum is very similar to the one measured on the black precipitate obtained from the solution of diphenylpolyynes in ref. [32]. The product of the polymerization of *sp*CCs terminated by aromatic groups can then be considered the intermediate step between the isolate *sp*CCs molecules and a solid made by *sp-sp$^2$* carbon [16,5]. Remarkably anyway, both the dried dinaphthylpolyynes and the precipitate do not show any decay evolution during their exposure to air (i.e., to oxygen). This indicates that the high reactivity of the *sp*CCs to oxygen observed in pure *sp-sp$^2$* carbon systems [16,40], is strongly inhibited by the presence of the hydrogen atoms in the aromatic terminal groups.

To gather further information on the stability of dried dinaphthylpolyynes, we repeated the process of vacuum drying of the solution at room temperature, and we recovered the dried dinaphthylpolyynes in the form of a dark orange powder. This powder was then characterized in a DSC, using a heating rate of 10°C/min under a $N_2$ flow. Figure 8 shows the resulting DSC trace. At 103.5°C can be observed a weak endothermic peak, indicating the onset of melting, probably for the shorter-chain oligomers. The peak is then followed by a broad exothermal transition with an onset temperature at 143°C and a peak at 182°C. The exotherm peak is quite broad, and the released energy is 442 J/g. This suggests the occurrence of a decomposition reaction for the dinaphthylpolyynes, probably initiated by the longer *sp*CCs (that are expected to be more unstable than the shorter ones [5,44]) and then propagating to the shorter ones.

**4. Conclusions**

In conclusion, we have shown how the Cadiot-Chodkiewicz reaction can be successfully applied to the synthesis of dinaphthylpolyynes. The key innovation in this synthetic approach is the use of diiodoacetylene, which simplifies the synthetic route to long polyyne chains with taylor-made end caps, thus indicating its general applicability in the synthesis of α,ω-diarylpolyynes.

HPLC analysis in conjunction with electronic absorption spectroscopy enabled us to identify the complete series of dinaphthylpolyynes with 2-6 acetylenic units. The length-resolved electronic spectroscopy of the dinaphthylpolyynes indicates, together with the well-known reduction of the HOMO-LUMO gap for increasing chain length, a nontrivial relation to the HOMO-LUMO gap of the end groups. This opens interesting perspectives in the interpretation of DIBs since *sp*CCs terminated by PAHs could give rise (if present in the interstellar medium) to peculiar absorption lines depending both to the chain length of the *sp*CC and to the structure of the PAHs.

By exploiting the observed independence of the molar exctinction coefficient of the $sp^2$-terminated *sp*CCs by the mass of the terminating organic groups, it was possible to estimate the relative molar concentration of the dinaphthylpolyynes produced. The obtained results indicate that the Cadiot-Chodkiewicz reaction favors the synthesis of longer chains when bulkier terminations are used.

UV-vis and FT-IR spectroscopy show that by evaporating the decalin solvent of the dinaphthylpolyynes solution, a solid film can be deposited where a relevant fraction of *sp*CCs is still present. Although a partial polymerization of the dinaphthylpolyynes has been observed, the obtained film is stable at room temperature even when it is exposed to air (i.e., to oxygen). By DSC, we checked the thermal stability of the dried dinaphthylpolyynes, observing the onset of their decomposition under nitrogen for temperatures higher than about 140°C. This moderately high thermal stability, together with the lack of reactivity at room temperature with oxygen, makes dinaphthylpolyynes promising building blocks for the integration of *sp*CCs in all-carbon electronic systems.

ACKNOWLEDGEMENTS

Franco Cataldo thanks the Italian Space Agency for the support of part of this research (Contract Number I/015/07/0 - Studi di Esplorazione del Sistema Solare). This work was supported by the European Union through the ETSF-I3 project (Grant Agreement No. 211956, user project 164).

Scheme 1. Chemical structures of the dinaphthylpolyynes synthesized with the Cadiot Chodkiewicz reaction between copper(I)-ethynylnaphthalide and diiodoacetylene.

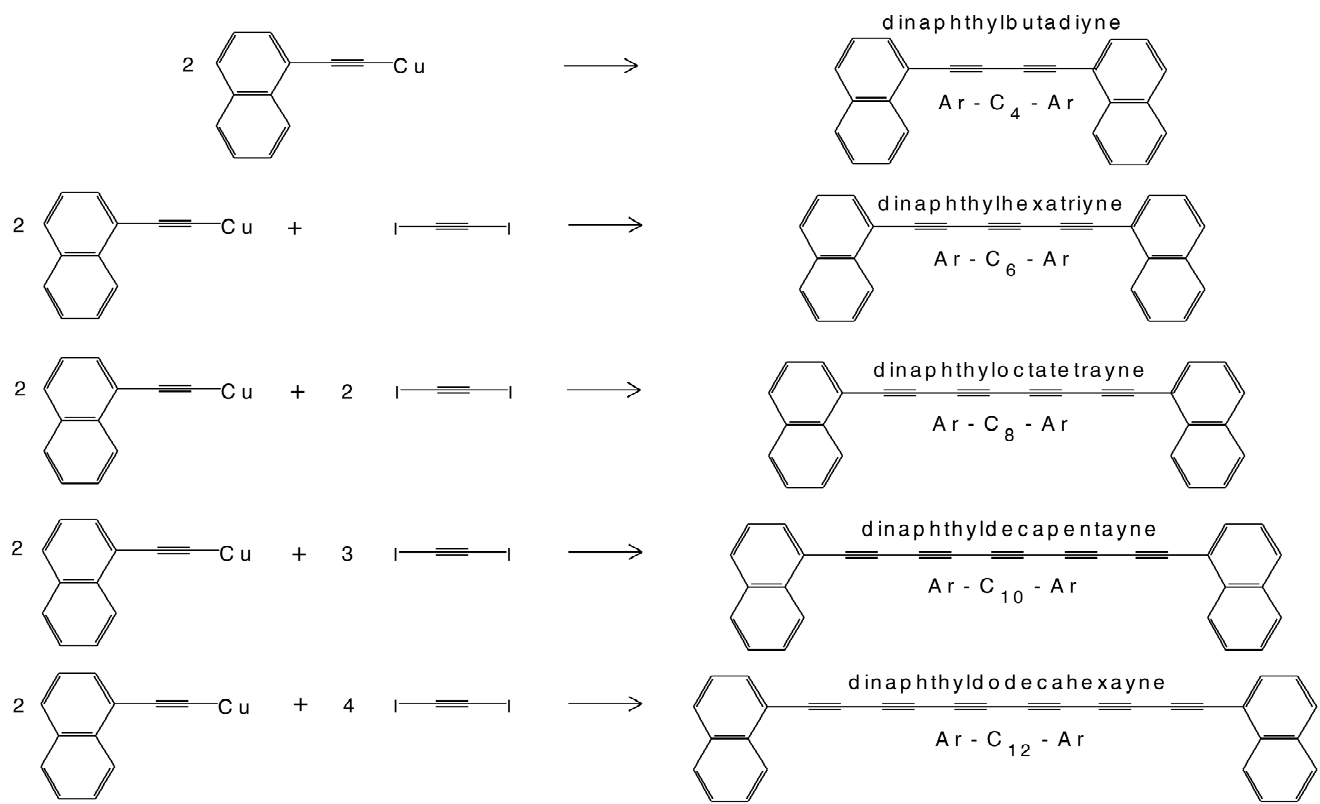

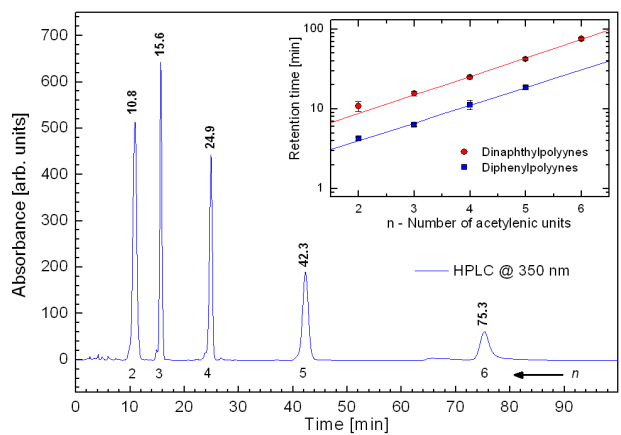

Figure 1. HPLC analysis of the dinaphthylpolyynes recorded at 350 nm. For each observed peak, the retention time and the corresponding number of acetylenic units (*n*) is indicated. Inset: comparison of the retention time of dinaphthylpolyynes and diphenylpolyynes [32] as a function of the number of acetylenic units componing the *sp*CCs. In both cases the mobile phase in the $C_8$ column was acetonitrile/water 80/20 vol./vol, and the pressure used was 105 bar.

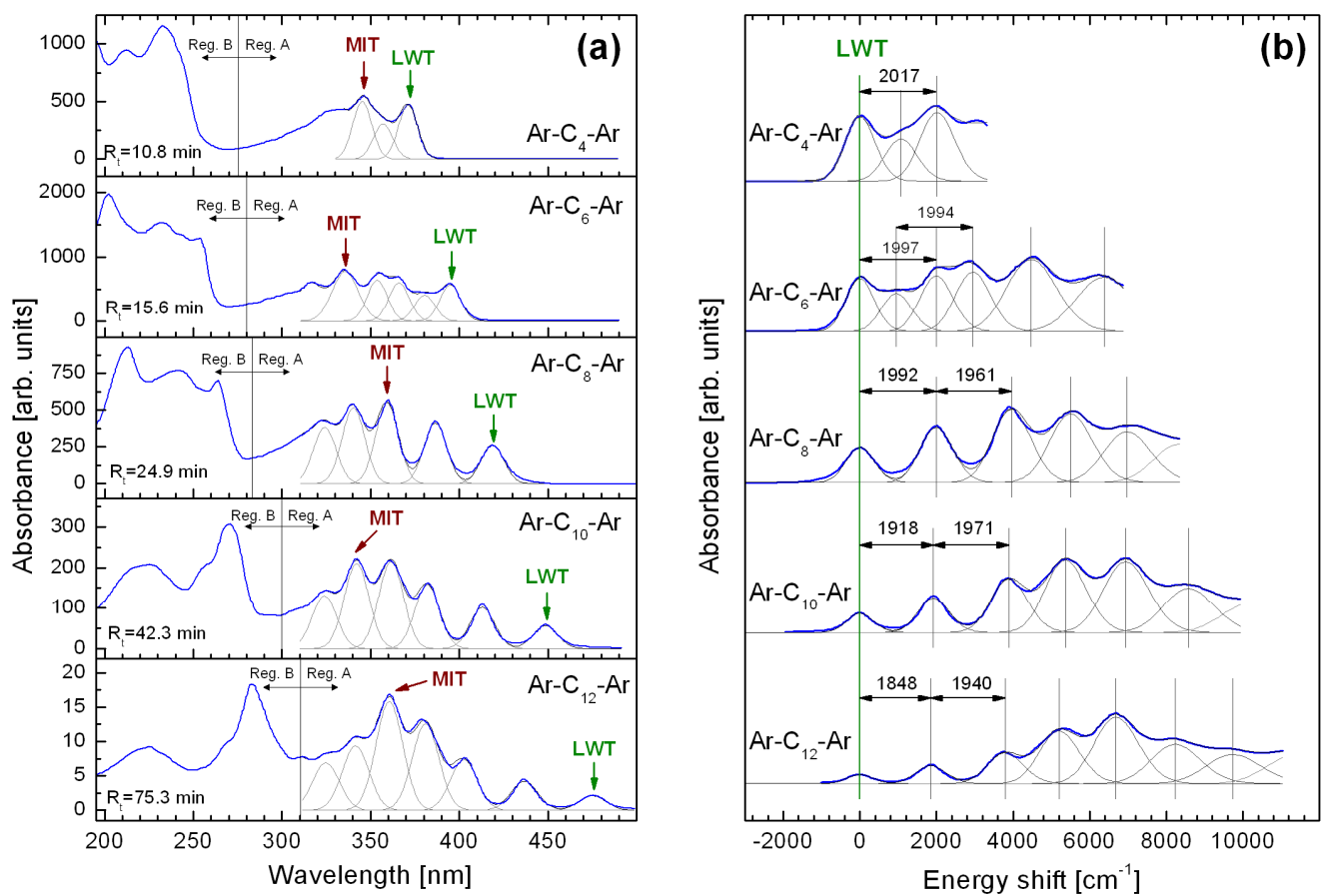

Figure 2. (a) The electronic absorption spectra of the dinaphthylpolyynes series (Ar-$C_{2n}$-Ar, $n$=2-6) as recorded by the diode array detector of the HPLC. For each spectrum the longest wavelength transition (LWT, inversely related to the HOMO-LUMO gap) and the most intense transition (MIT) are indicated. (b) The electronic absorption spectra of the dinaphthylpolyynes series (Ar-$C_{2n}$-Ar, $n$=2-6) represented as a function of the energy shift from the LWT (expressed in cm$^{-1}$). The distance between the peaks in the absorption spectra is consistent with the phonon energy of the optical C-C stretching mode of the *sp*-carbon chain.

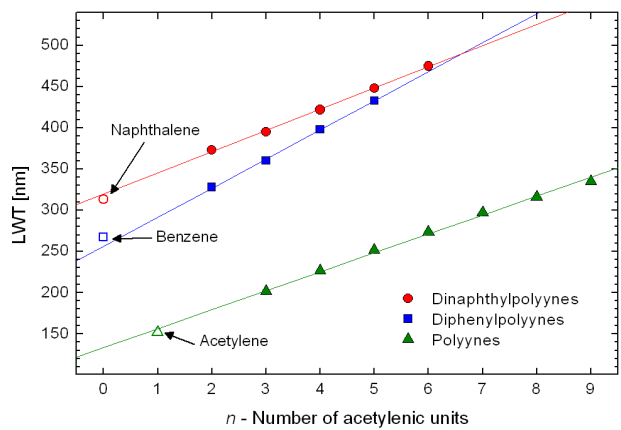

Figure 3. The position of the LWT as a function of the number of acetylenic units composing the carbon chain of three *sp*CCs series: naphthyl-terminated (dinaphthylpolyynes, circles), phenyl-terminates (diphenylpolyynes, squares) [32] and hydrogen-terminated (polyynes, triangles) [2]. The LWT for benzene and naphthalene molecules are indicated at *n*=0.

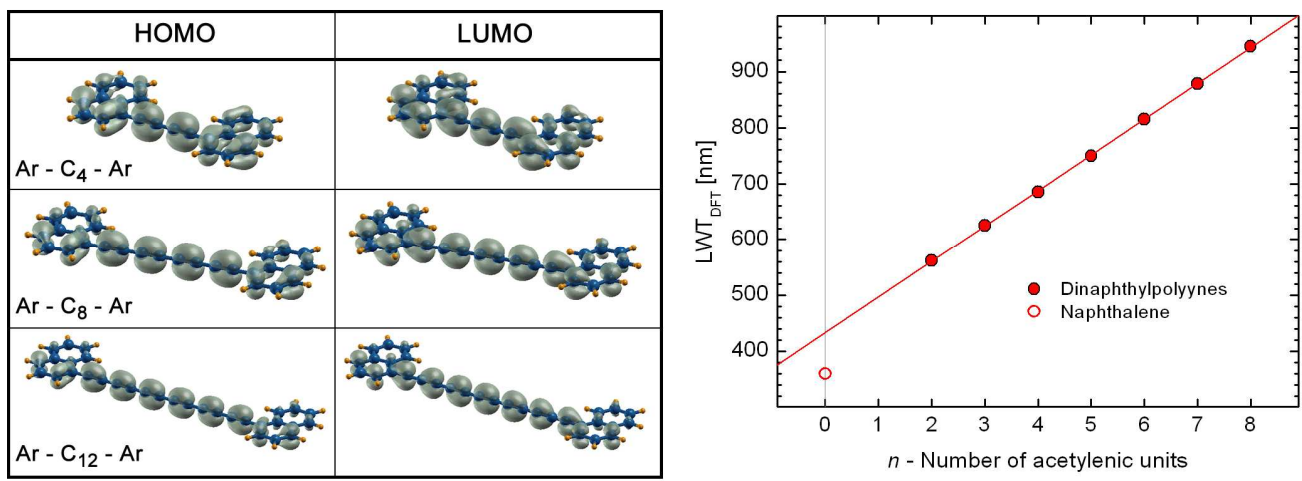

Figure 4. (Left panel) The electron density of HOMO and LUMO states for three different dinaphthylpolyynes: Ar-$C_4$-Ar, Ar-$C_8$-Ar, Ar-$C_{12}$-Ar. As the *sp*-carbon chain is longer, the HOMO and LUMO states become more and more localized on the chain backbone. (Right panel) DFT-LSDA value of LWT$_{DFT}$ (evaluated as the inverse of the HOMO-LUMO Kohn-Sham gap) as a function of the number of acetylenic units composing the carbon chain (*n*) for dinaphthylpolyynes. The evaluated LWT$_{DFT}$ for a single naphthalene molecule is indicated at *n*=0.

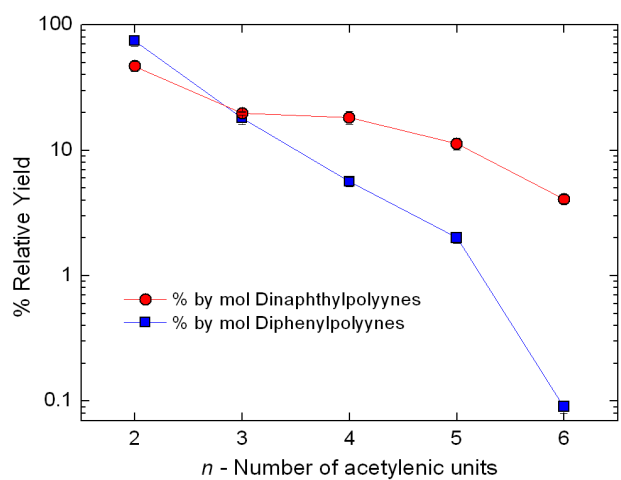

Figure 5. Relative yield of dinaphthylpolyynes and diphenylpolyynes [32] as a function of the number of acetylenic units (*n*). Both series of *sp*CCs are produced by the Cadiot-Chodkiewcz reaction.

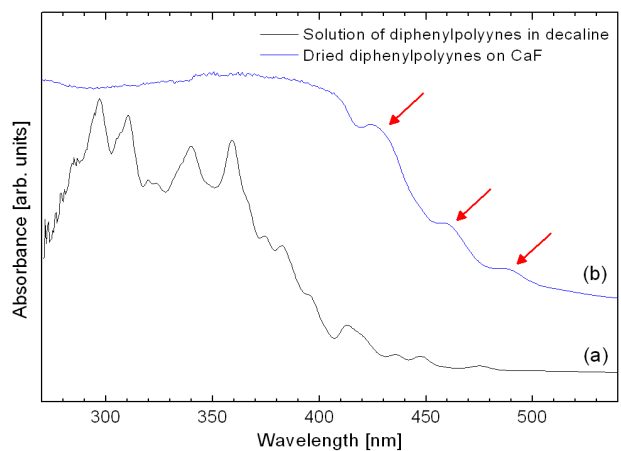

Figure 6. The UVvis spectra of the dinaphthylpolyynes solution (a) and of the dinaphthylpolyynes solution dried on a CaF substrate (b). Spectrum (a) is consistent with the weighted sum of the spectra of the separated dinaphthylpolyynes in Figure 2.

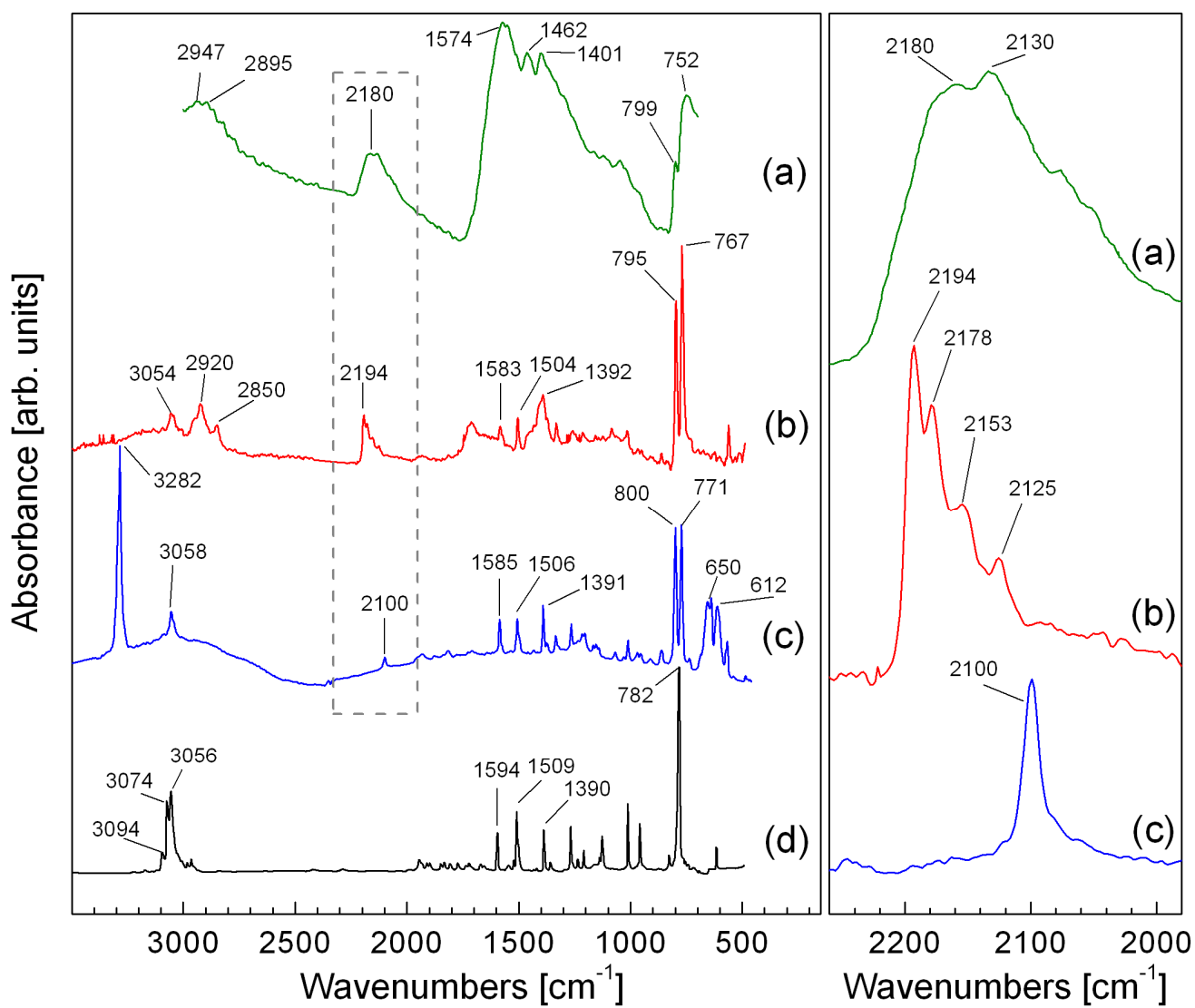

Figure 7. (Left) The FT-IR spectra of the black precipitate produced during storage at room temperature (in closed flasks) of the dinaphthylpolyynes solution (a), of the dried dinaphthylpolyynes solution (b), of pure 1-ethynylnaphthalene (c) and of pure naphthalene (d) (from Ref. [41]). Spectrum (a) has been acquired in reflectance mode depositing the sample on ZnSe, while spectrum (b) has been acquired in transmittance mode after drying the solution on a KBr substrate. (Right) Detail of the triple bonds stretching region of the FT-IR spectra (a), (b) and (c). A linear background has been subtracted to spectra (a) and (c).

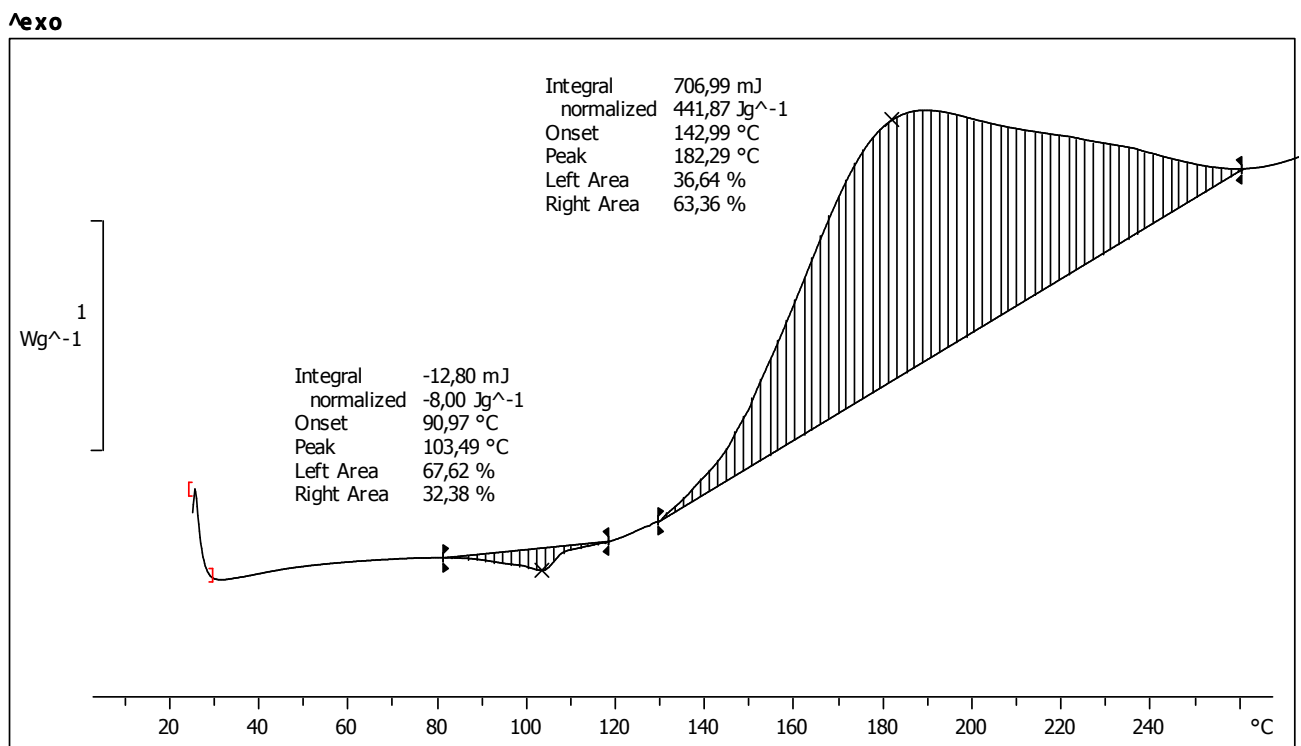

Figure 8. Differential scanning calorimetry (DSC), 10°C/min under $N_2$, of the dinaphtylpolyynes mixture. The exothermic reaction with onset at 143°C and peak at 182°C can be assigned to the thermal decomposition of the dinaphthylpolyynes.

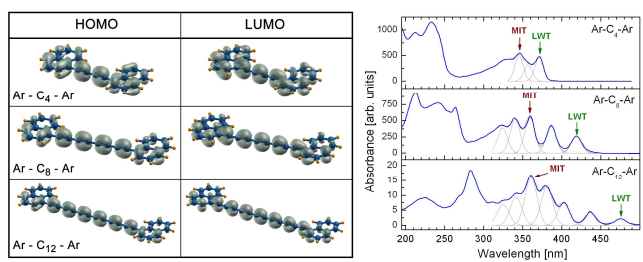

(Left) The electron density of HOMO and LUMO states evaluated by DFT-LSDA simulations for three different dinaphthylpolyynes: Ar-$C_4$-Ar, Ar-$C_8$-Ar, Ar-$C_{12}$-Ar. As the *sp*-carbon chain is longer, the HOMO and LUMO states become more and more localized on the chain backbone. (Right) The experimental electronic absorption spectra of Ar-$C_4$-Ar, Ar-$C_8$-Ar and Ar-$C_{12}$-Ar as recorded by the diode array detector of the HPLC. The observed evolution of the longest wavelength transition (LWT, inversely related to the HOMO-LUMO gap) is in good agreement with the theoretical predictions.